\documentclass[vecphys]{svmult}

\usepackage{makeidx}         
\usepackage{graphicx}        
\usepackage{multicol}        
\usepackage[bottom]{footmisc}

\usepackage{amssymb}

\makeindex             

\begin{document}

\title*{The QUIJOTE CMB Experiment}
\author{J.A. Rubi\~no-Mart\'in\inst{1}\and
R.~Rebolo\inst{1}\and
M.~Tucci\inst{1}\and
R.~G\'enova-Santos\inst{1}\and
S.R.~Hildebrandt\inst{1}\and
R.~Hoyland\inst{1}\and
J.M.~Herreros\inst{1}\and
F.~G\'omez-Re\~nasco\inst{1}\and
C.~L\'opez Caraballo\inst{1}\and
E.~Mart\'inez-Gonz\'alez\inst{2}\and
P.~Vielva\inst{2}\and
D.~Herranz\inst{2}\and
F.J.~Casas\inst{2}\and
E.~Artal\inst{3}\and
B.~Aja\inst{3}\and
L.~de la Fuente\inst{3}\and
J.L.~Cano\inst{3}\and
E.~Villa\inst{3}\and
A.~Mediavilla\inst{3}\and
J.P.~Pascual\inst{3}\and
L.~Piccirillo\inst{4}\and
B.~Maffei\inst{4}\and
G.~Pisano\inst{4}\and
R.A.~Watson\inst{4}\and
R.~Davis\inst{4}\and
R.~Davies\inst{4}\and
R.~Battye\inst{4}\and
R.~Saunders\inst{5}\and
K.~Grainge\inst{5}\and
P.~Scott\inst{5}\and
M.~Hobson\inst{5}\and
A.~Lasenby\inst{5}\and
G.~Murga\inst{6}\and
C.~G\'omez\inst{6}\and
A.~G\'omez\inst{6}\and
J.~Ari\~no\inst{6}\and
R.~Sanquirce\inst{6}\and
J.~Pan\inst{6}\and
A.~Vizcarg\"uenaga\inst{6}\and
B.~Etxeita\inst{6}}
\authorrunning{J.A. Rubi\~no-Mart\'in et al. }
\institute{
Instituto de Astrofisica de Canarias (IAC),
C/Via Lactea, s/n, E-38200, La Laguna, Tenerife, Spain.
Corresponding author: \texttt{jalberto@iac.es}
\and
Instituto de Fisica de Cantabria (IFCA), CSIC-Univ. de Cantabria, Avda. los
Castros, s/n, E-39005 Santander, Spain.
\and
Departamento de Ingenieria de COMunicaciones (DICOM),
Laboratorios de I+D de Telecomunicaciones, Plaza de la Ciencia s/n, E-39005
Santander, Spain.
\and
Jodrell Bank Centre for Astrophysics, School of Physics and Astronomy,
University of Manchester, Oxford Road, Manchester M13 9PL, UK
\and
Astrophysics Group, Cavendish Laboratory, University of Cambridge, Madingley
Road, Cambridge CB3 0HE
\and
IDOM, Avda. Lehendakari Aguirre, 3, E-48014 Bilbao, Spain.
}
\maketitle
\index{J.A. Rubi\~no-Martin}

\begin{abstract}
We present the current status of the QUIJOTE (Q-U-I JOint TEnerife) CMB
Experiment, a new instrument which will start operations early 2009 at Teide
Observatory, with the aim of characterizing the polarization of the CMB and
other processes of galactic and extragalactic emission in the frequency range
10-30~GHz and at large angular scales. QUIJOTE will be a valuable complement at
low frequencies for the PLANCK mission, and will have the required sensitivity
to detect a primordial gravitational-wave component if the tensor-to-scalar
ratio is larger than $r=0.05$.
\end{abstract}

\section{Introduction}

The study of the Cosmic Microwave Background (CMB) anisotropies is one of the
main pillars of the Big Bang model. With the latest results from WMAP satellite
\cite{WMAP5}, and the information provided by ground-based experiments such as
VSA \cite{VSA}, ACBAR \cite{ACBAR} or CBI \cite{CBI}, it has been possible to
determine cosmological parameters with accuracies better than 5\% (see
e.g. \cite{Dunkley2008}).

However, the CMB contains far more information encoded in its polarization
signal. Since the first detection of polarization by the DASI
experiment\cite{DASIpol}, other experiments have started to measure the angular
power spectrum of the polarization.  Although those measurements are still
having a relatively poor signal-to-noise, they show excellent agreement with the
predictions of the standard $\Lambda$CDM model.

The standard theory predicts that the CMB is linearly polarized, the physical
mechanism responsible for its polarization being Thomson scattering during the
recombination or reionization epochs.
Generally speaking, the polarization tensor can be decomposed in terms of a
E-field (gradient) and a B-field (rotational) components
\cite{ZaldaSeljak07,Kamion97}.
Due to parity conservation, this implies that we have three angular power
spectra to describe the polarization field: the TE (cross-correlation of
temperature and E mode), the EE and BB power spectra. All the other combinations
(TB and EB) should be zero.

If the fluctuations in CMB intensity are seeded by scalar perturbations (i.e
fluctuations in the density alone), one would only expect primordial E modes in
the CMB polarization. However, vector and tensor perturbations, like those due
to gravitational waves in the primordial Universe (e.g. \cite{Polnarev85}), are
mechanisms that could generate primordial B-modes in the polarization on large
angular scales.
Therefore if we can measure these modes we may have a unique way to carry out a
detailed study of the inflationary epoch. In particular, the energy scale $V$ at
which inflation occurred can be expressed in terms of $r$, the ratio of tensor to
scalar contributions to the power spectrum, as \cite{Partridge}
\begin{equation}
r = 0.001 \Bigg( \frac{V}{ 10^{16}~{\rm GeV}} \Bigg)^4
\end{equation}
The current upper limit of $r \lesssim 0.3$ from WMAP data \cite{Komatsu08}
translates into $V \lesssim 4 \times 10^{16}$~GeV.

Because of the importance of detecting primordial gravitational waves
\cite{TaskForce,ESAESO}, there is a huge interest to develop ground-based
experiments to measure (or constrain) the amplitude of B-modes power spectrum of
the CMB polarization. Here we present one of these efforts.

The QUIJOTE (Q-U-I JOint TEnerife) CMB Experiment is a scientific collaboration
between the Instituto de Astrof\'isica de Canarias, the Instituto de F\'isica de
Cantabria, the IDOM company, and the universities of Cantabria,
Manchester and Cambridge, 
with the aim of characterizing the polarization of the CMB, and other galactic
and extragalactic physical processes in the frequency range 10-30~GHz and at
angular scales larger than 1 degree. Updated information of the project can be
found at the following web page: \verb!http://www.iac.es/project/cmb/quijote!.

\begin{figure}
\centering
\includegraphics[height=5.5cm]{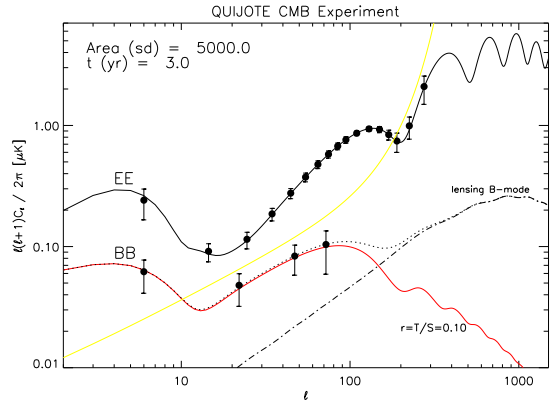}
\caption{QUIJOTE scientific goal for the angular power spectrum of the CMB E and
  B mode signals. It is shown the case for 3~years operation time, and a sky
  coverage of $\sim 5,000$~square degrees. The red line corresponds to the
  primordial B-mode contribution in the case of $r=0.1$. Yellow line indicates
  the associated QUIJOTE noise power spectrum. Dots with error bars correspond
  to averaged measurements over a certain multipole band. }
\label{fig:goal}
\end{figure}

\section{Science goals}

The QUIJOTE-CMB experiment has two primary scientific goals:
\begin{itemize}
\item to detect the imprint of gravitational B-modes if they have an amplitude
$r \ge 0.05$;
\item to provide essential information of the polarization of the synchrotron and
  the anomalous microwave emissions from our Galaxy at low frequencies
  (10-20~GHz).
\end{itemize}

In order to achieve these scientific objectives, we need to cover a total sky
area of the order of 3,000 - 10,000 square degrees, and to reach sensitivities
of $\sim 3-4$~$\mu$K per one degree beam after one year of operation with the
low frequency instrument (11-19~GHz), and $\lesssim 1$~$\mu$K per beam with the
second instrument at 30~GHz. Although the final observing strategy is still
under discussion, a possible solution is presented in Fig.~\ref{fig:goal}, where
we show the scientific goal for the angular power spectrum of the E and B modes
after 3 years of operation, assuming a sky coverage of 5,000 square degrees. In
this particular case, the final noise figure for the 30~GHz map is $\sim
0.5$~$\mu$K/beam.

\begin{figure}
\centering
\includegraphics[height=5cm]{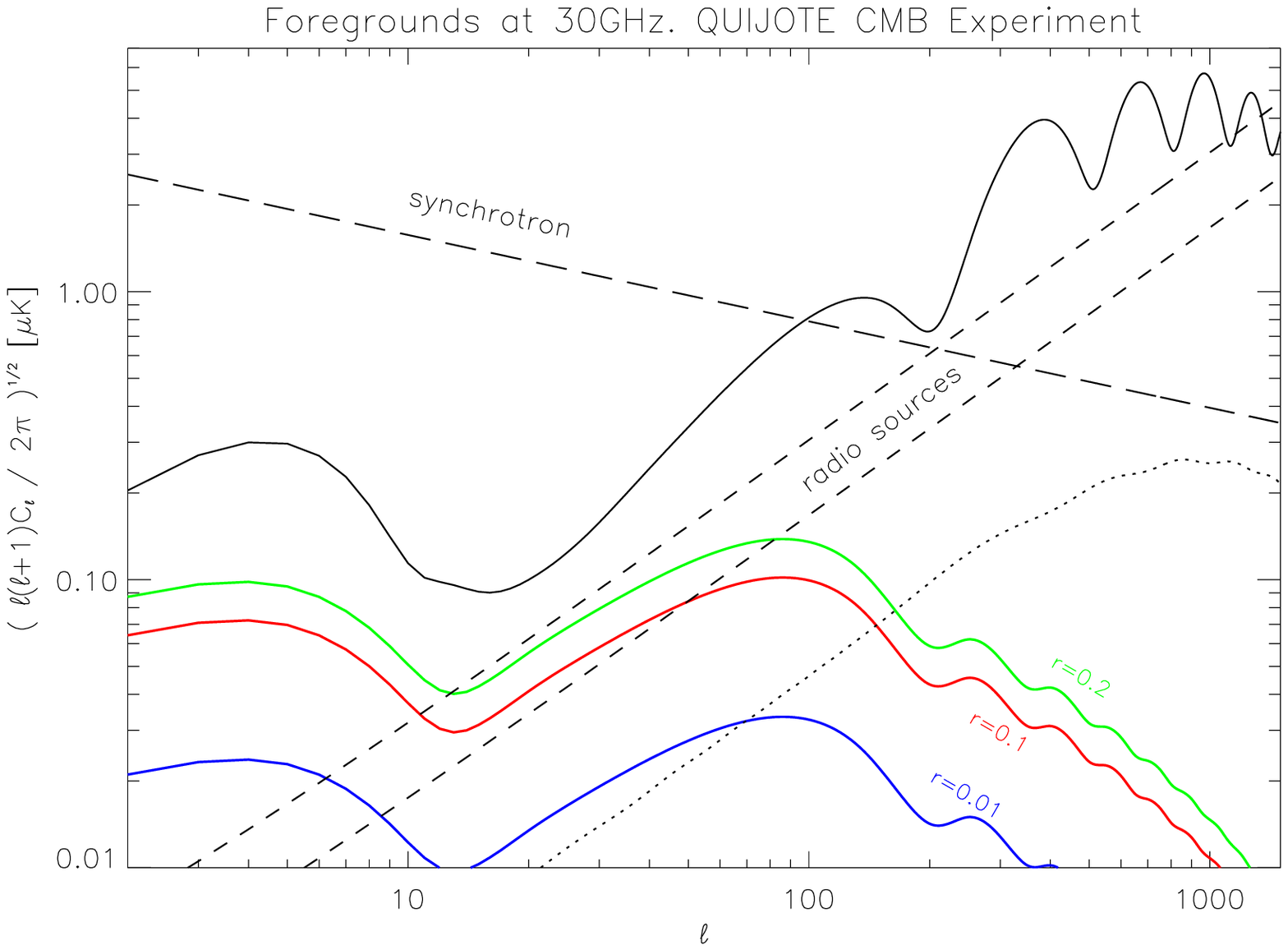}%
\includegraphics[height=5cm]{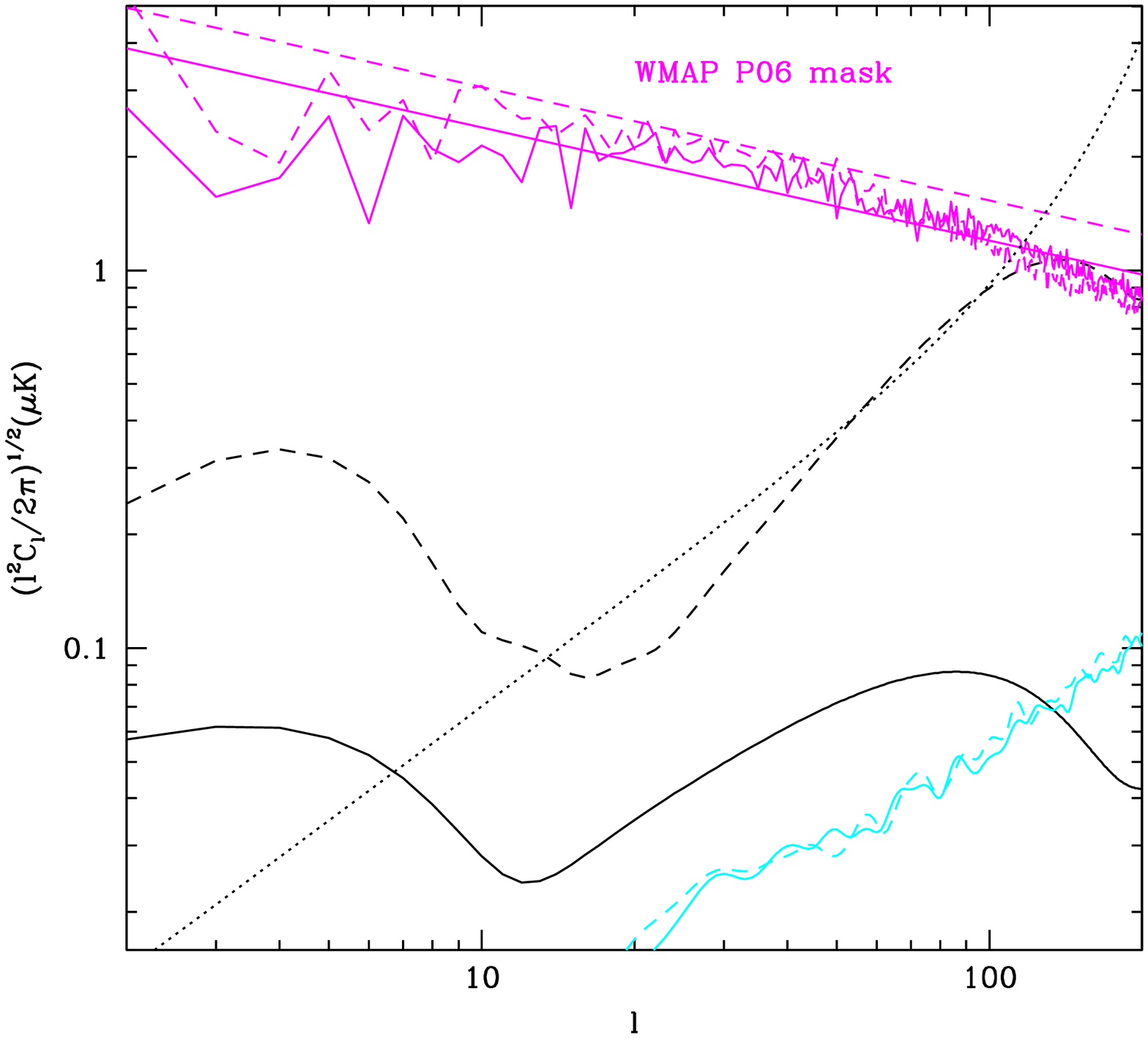}
\caption{Left: Expected foreground contamination in the 30~GHz QUIJOTE frequency
  band. It is shown the contribution of polarized synchrotron emission and
  radio-sources for the case of subtracting sources down to 1~Jy in total
  intensity (upper dashed line for radio-sources) and 300~mJy (lower
  dashed-line). Right: Correction of the synchrotron emission at 30~GHz. The
  magenta line shows the predicted synchrotron level from a simulation of the
  expected signal. Assuming a pure power-law dependence for the synchrotron
  emission in the QUIJOTE frequency range, the blue line shows the synchrotron
  residual after correction using the low frequency channels (11 to 19~GHz). }
\label{fig:foregrounds}
\end{figure}

According to those nominal sensitivities, QUIJOTE will provide one of the most
sensitive 11-19~GHz measurements of the polarization of the synchrotron and
anomalous emissions on degree angular scales.
This information is extremely important given that B-modes are known to be
sub-dominant in amplitude as compared to the Galactic emission (see
e.g. \cite{Tucci}). This is illustrated in the left panel of
Fig.~\ref{fig:foregrounds}, where we present the amplitude of the expected
synchrotron and radio-source contribution at 30~GHz, computed according to the
models described in \cite{Tucci}.

The QUIJOTE low frequency maps will complement the measurements of the Planck
satellite\footnote{PLANCK: http://www.rssd.esa.int/index.php?project=Planck},
helping in the characterization of the Galactic emission. In particular, QUIJOTE
will provide a key contribution to assess the level of a possible contribution
of polarized microwave anomalous emission \cite{Watson,Battistelli}.

Using the low frequency maps, we plan to correct the high frequency QUIJOTE
channel (30~GHz) to search for primordial B-modes. To illustrate this,
Fig.~\ref{fig:foregrounds} shows the residual synchrotron contribution after a
pixel-by-pixel correction of the high frequency map assuming a pure power-law
behavior for the synchrotron emission law.
The issue of radio-sources is discussed below, in the context of the source
subtraction facility.

\section{Experimental details}

\subsection{Project baseline}

QUIJOTE-CMB will observe at five frequencies, namely 11, 13, 17, 19 and 30~GHz,
with an angular resolution of $\sim 1$~degree. It will operate from the
Observatorio del Teide (2400~m) in Tenerife (Spain), which has been shown to be
an excellent place for CMB observations (same site as Tenerife, COSMOSOMAS, VSA
and JBO-IAC Interferometer experiments).

The project has two phases.
\textit{Phase I}, which is completely funded, consists in the construction of a
first telescope and two instruments which can be exchanged in the focal
plane. The first instrument will be a multichannel instrument providing the
frequency coverage between 11 and 19~GHz, plus a single pixel at 30~GHz, and it
is expected to start observations at the beginning of 2009. The second
instrument will consist of 19 polarimeters working at 30~GHz, and it is expected
to start operations at the end of 2009.
Table~\ref{tab:basic} summarizes the basic parameters describing these two
instruments\footnote{Note that our definition for Stokes parameters is such that
$Q=T_{\rm x} - T_{\rm y}$. }. The temperature sensitivity per beam is computed
as
\begin{equation}
\Delta Q = \Delta U = \sqrt{2} \frac{ T_{\rm sys} }{ \sqrt{\Delta \nu \; t \;
    N_{chan}} },
\end{equation}
being $N_{\rm chan}$ the number of channels, $\Delta \nu$ the bandwidth and
$T_{\rm sys}$ the system temperature (i.e. including the sky). \textit{Phase I}
also includes a source subtractor facility to monitor and correct the
contribution of polarized radio-sources in the final maps. The overall time
baseline for the project is to achieve the main science goal ($r=0.1$) by end of
2011, and $r=0.05$ by 2015.

Finally, \textit{Phase II} (which is still not funded) considers the
construction of a second telescope identical to the first one, and a third
instrument with 30 polarimeters at 40~GHz.

\begin{table}
\centering
\caption{QUIJOTE-CMB Experiment. Instruments characteristics. }
\label{tab:basic}
\begin{tabular}{l@{ \hspace{0.1cm} }c@{ \hspace{0.1cm} }c@{ \hspace{0.1cm} }c@{ \hspace{0.1cm} }c@{ \hspace{0.1cm} }c@{ \hspace{0.2cm} }c}
\hline\noalign{\smallskip}
 & \multicolumn{5}{c}{Instrument I} & Instr. II \\
\cline{2-7}
Frequency [GHz]              &     11.0 &  13.0 &  17.0 &  19.0 &  30.0 &  30.0 \\
Bandwidth [GHz]              &      2.0 &   2.0 &   2.0 &   2.0 &   8.0 &   8.0 \\
Number of channels           &        8 &     8 &     8 &     8 &     2 &    38 \\
Beam FWHM [deg]              &     0.92 &  0.92 &  0.60 &  0.60 &  0.37 &  0.37 \\
Tsys [K]                     &     20.0 &  20.0 &  20.0 &  20.0 &  30.0 &  20.0 \\
Sensitivity [Jy s$^{1/2}$]   &     0.24 &  0.34 &  0.24 &  0.30 &  0.43 &  0.07 \\
Sens per beam [mK $s^{1/2}$] &     0.22 &  0.22 &  0.22 &  0.22 &  0.34 &  0.05 \\
\noalign{\smallskip}\hline
\end{tabular}
\end{table}

\subsection{Telescope and enclosure}

\begin{figure}
\centering
\includegraphics[width=10cm]{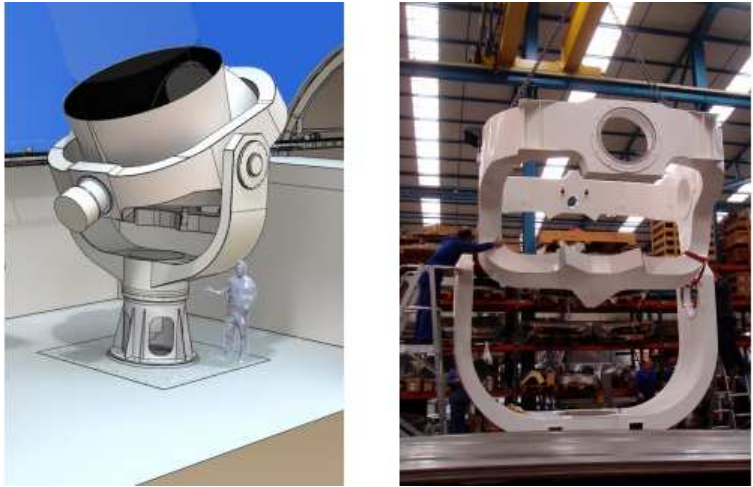}
\caption{Left: Design of one of the QUIJOTE telescopes inside the
  enclosure. Right: Assembling the first QUIJOTE telescope (October 7th,
  2008). }
\label{fig:tel}
\end{figure}

The QUIJOTE-CMB telescope uses a crossed-Dragonian design, where the primary has
a 3~m projected aperture, and the secondary 2.6~m. The system is
under-illuminated to minimize sidelobes and ground spillover. In addition, a
cylindrical absorbing screen surrounding the optics (see Fig.~\ref{fig:tel})
minimizes the spillover signal.
Both mirrors have been designed to operate up to 90~GHz i.e. rms $\le 20$~$\mu$m
and maximum deviation of $d=100$~$\mu$m.

The whole system is mounted on a platform that can rotate around the vertical
axis at a frequency of 0.25~Hz.
The supporting structure has been designed using an alto-azimuthal concept which
enables the telescope to point to any position in the sky with elevation above
the horizon higher than $30^\circ$.

\subsection{First instrument}

\begin{figure}
\centering
\includegraphics[height=6cm]{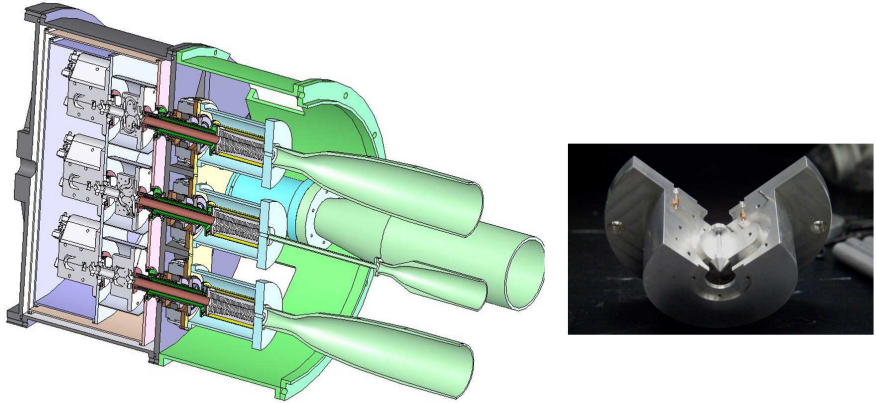}
\caption{Left: Detailed design of the first instrument. It is shown four horns
  (the 30~GHz is the central one, two horns at 17-19~GHz, and one of the
  11-13~GHz horns in the back). After each antenna, we see the waveguide, the
  polar modulator and the OMT.  Right: A prototype of the final polar
  modulator. }
\label{fig:first}
\end{figure}

This is a multi-channel instrument with five separate polarimeters (providing 5
independent sky pixels): two which operate at 10-14~GHz, two which operate at
16-20~GHz, and a central polarimeter at 30~GHz.
The science driver for this first instrument is the characterization of the
galactic emission. The optical arrangement includes 5 conical corrugated
feedhorns (designed by the University of Manchester) staring into a dual
reflector crossed-dragonian system, which provides optimal cross-polarization
properties (designed to be $\le -35$~dB) and symmetric beams.
Each horn feeds a novel cryogenic on-axis rotating polar modulator which can
rotate at speeds of up to 40~Hz (see Fig.~\ref{fig:first}). This rotational rate
is fast enough to switch out 1/f noise in the lower frequency LNAs (since polar
modulation occurs at four times the rotational rate, i.e. $160$~Hz).
The 30~GHz Front-End module (FEM) has additional phase switching to provide
stability. The orthogonal linear polar signals are separated through a wide-band
cryogenic Ortho-Mode-Transducer (OMT) before being amplified through two similar
LNAs (a Faraday type module in the case of 30~GHz).
These two orthogonal signals are fed into a room-temperature Back-End module
(BEM) where they are further amplified and spectrally filtered before being
detected by square-law detectors.
All the polarimeters except the 30~GHz receiver have simultaneous "Q" and "U"
detection i.e. the 2 orthogonal linear polar signals are also correlated through
a $180^\circ$ hybrid and passed through two additional detectors. The band
passes of these lower frequency polarimeters have also been split into an upper
and lower band which gives a total of 8 channels per polarimeter (see
Table~\ref{tab:basic}).

The FEM for the low frequency channels is being built by IAC. The receivers for
these channels use MMIC 6-20~GHz LNAs (designed by S. Weinreb and built in
Caltech). The gain for these amplifiers is approximately 30~dB, and the noise
temperature is less than 9~K across the band.
The 30~GHz FEM is being built at the University of Manchester, and the design
uses an existing Faraday module (same as the one used for
OCRA-F\footnote{OCRA-F:
http://www.jodrellbank.manchester.ac.uk/research/ocra/ocraf.html.}).
The BEM for the 30~GHz instrument is being built by DICOM, with collaboration of
IFCA at the simulation level.
The cryogenics and the mechanical systems are provided by CMS\footnote{CMS:
http://www.cryo-mechanicalsystems.com/} (Jeff Julian), IDOM and IAC.

\subsection{Second instrument}

This instrument will be devoted to primordial B-modes science. It will consist
in 19 polarimeters operating at 30~GHz. The conceptual design is a re-scaled
version of the first instrument.

\subsection{Source subtractor facility}

\begin{figure}
\centering
\includegraphics[height=5cm]{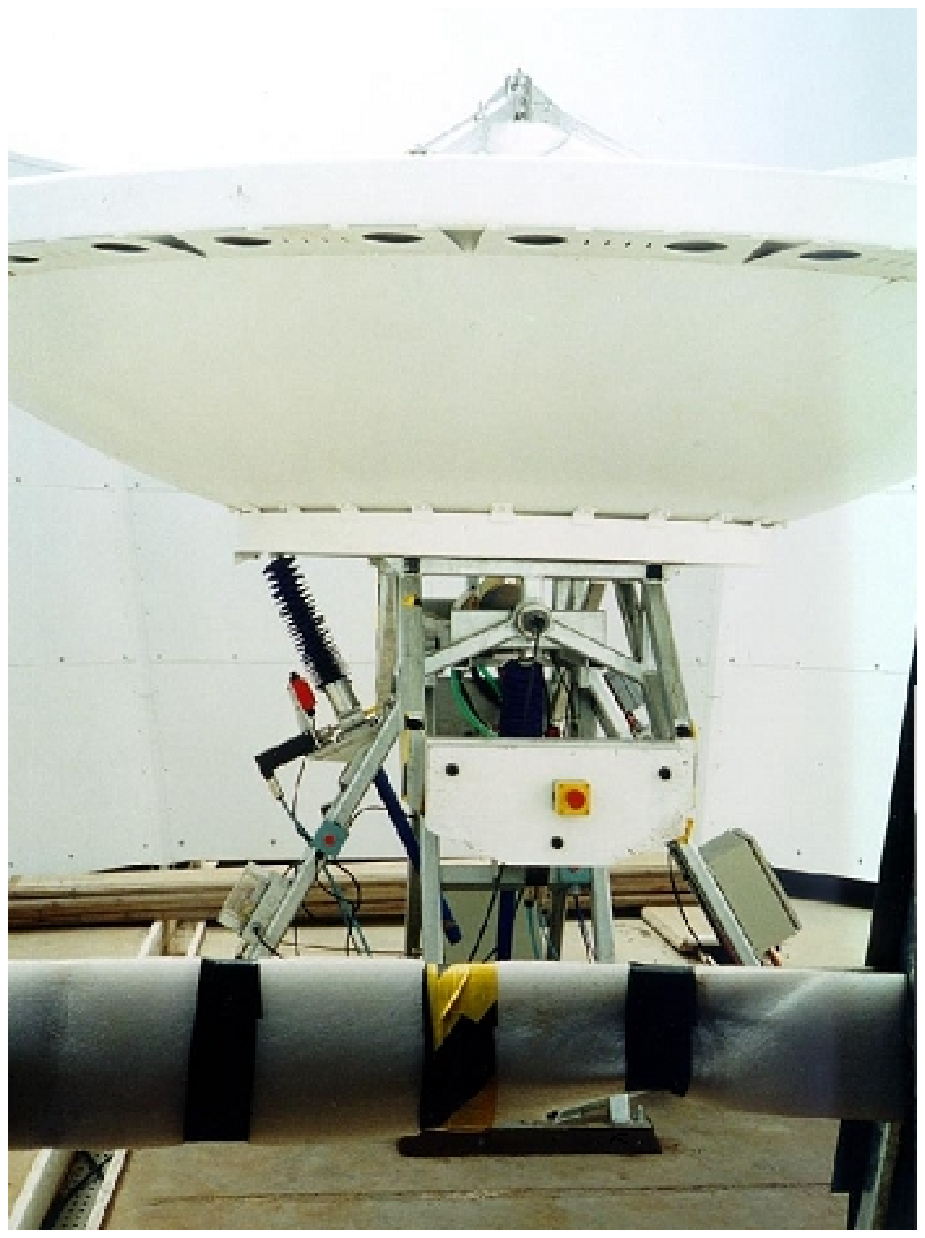}%
\includegraphics[height=5cm]{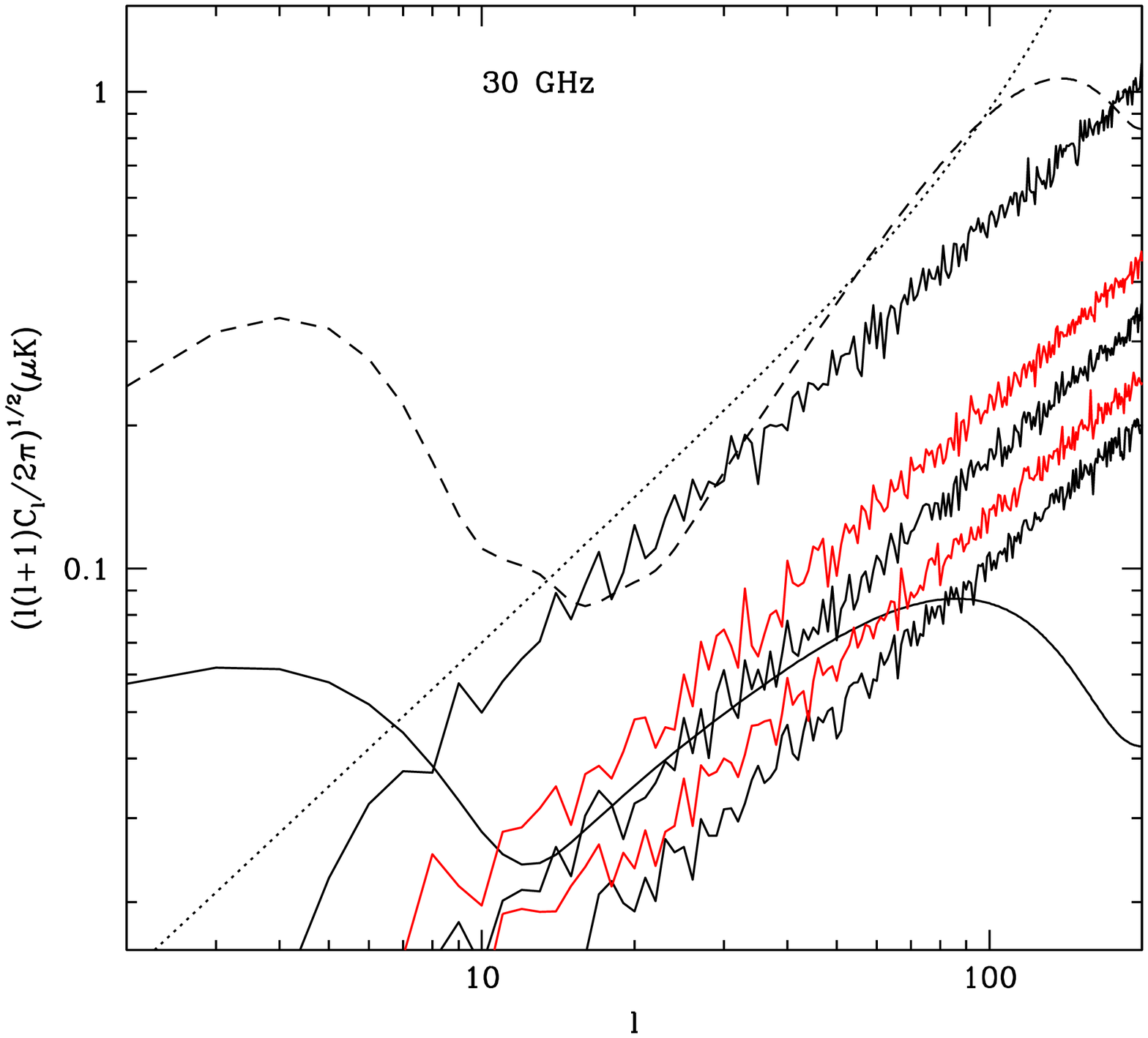}
\caption{Left: One antenna of the VSA source subtractor. This facility will be
  re-used to measure polarization of radio sources to correct the QUIJOTE 30~GHz
  maps. Right: Radio-source contribution at 30~GHz (top solid oscillating line),
  and the residuals after removing/masking sources with fluxes (in total
  intensity) higher than 300~mJy or 100~mJy (black lines). Solid red lines are
  obtained removing sources with polarized intensity higher than 50 or 10~mJy. }
\label{fig:vsa-ss}
\end{figure}

An upgraded version of the VSA source subtractor (VSA-SS) facility \cite{VSA1},
which is being carried out by the Cavendish Laboratory and the University of
Manchester, will be used to monitor the contribution of radio-sources in the
QUIJOTE maps. The VSA-SS is a two element interferometer, operating at 30~GHz,
with $3.7$~m dishes and a separation of 9~m (see Fig.~\ref{fig:vsa-ss}).
We have estimated that at 30~GHz it is enough to correct the emission of all
sources with fluxes in total intensity higher than 300~mJy in order to make the
residual source contribution equal or smaller than the expected B-mode signal
for the case of $r=0.1$ (see Fig.~\ref{fig:vsa-ss}). In that case, the total
number of sources to be monitored in the whole QUIJOTE surveyed area is around
$500$. The expected flux sensitivity per source is $\sim 2-3$~mJy.
%

\section{Conclusions}

QUIJOTE-CMB will provide unique information about the polarization emission
(synchrotron and anomalous) from our Galaxy at low frequencies. This information
will be valuable for future B-mode experiments. In particular, QUIJOTE will
complement at low frequencies the information obtained by Planck.
Using the information from the low frequencies, QUIJOTE will be able to detect
the B-mode signal due to primordial gravity waves in the 30~GHz map if $r \ge
0.05$.

%
%
%


\printindex
\end{document}